\documentclass{mem}
\usepackage{natbib}
\usepackage{txfonts}
\usepackage{graphicx}
\usepackage{epstopdf}		% To use .eps in TexWorks
\usepackage[a4paper,breaklinks,dvipdfm]{hyperref}
\usepackage{url}			% For breaking URLs easily trough lines
\usepackage{subfigure}
\usepackage{balance}
\newcommand{\COBOLD}{{\tt CO$^5$BOLD}}
\newcommand{\Stagger}{{\tt Stagger}}

\newcommand{\LINFOR}{{\tt Linfor3D}}

\newcommand{\moh}{\ensuremath{[\mathrm{M/H}]}}
\newcommand{\feoh}{\ensuremath{[\mathrm{Fe/H}]}}
\newcommand{\Teff}{\ensuremath{T_{\mathrm{eff}}}}

\def\fei{\ion{Fe}{i}}

\def\vsini{$v \sin{i}$}

\begin{document}

	\title{Spectral line asymmetries in the metal-poor red giant HD~122563:\\
    \COBOLD\ predictions versus observations}
	\titlerunning{Spectral line asymmetries in the metal-poor red giant HD~122563}
	\authorrunning{Klevas et al.}

	\author{J. Klevas\inst{1},
	A. Ku\v{c}inskas\inst{1,2}
	H.-G. Ludwig,  \inst{3}
	P. Bonifacio,  \inst{4}
	M. Steffen,    \inst{5}
	\and D.~Prakapavi\v{c}ius  \inst{1}
	}

	\offprints{J.~Klevas}

	\institute{Vilnius University Institute of Theoretical Physics and Astronomy,
    Go\v{s}tauto 12, Vilnius LT-01108, Lithuania; \email{jonas.klevas@tfai.vu.lt}
	\and
	Vilnius University Astronomical Observatory, \v{C}iurlionio 29, Vilnius
    LT-03100, Lithuania
	\and
	ZAH Landessternwarte K\"{o}nigstuhl, D-69117 Heidelberg, Germany
	\and
	GEPI, Observatoire de Paris, CNRS, Universit\'{e} Paris Diderot, Place Jules
    Janssen, 92190 Meudon, France
	\and
	Leibniz-Institut f\"ur Astrophysik Potsdam, An der Sternwarte 16, D-14482
    Potsdam, Germany
	}

    \abstract{
    We study the influence of convection on the asymmetries and Doppler shifts of \ion{Fe}{i} spectral lines in the metal-poor red giant HD~122563. To this end, we compute theoretical \ion{Fe}{i} line shifts and line bisectors using 3D hydrodynamical model atmosphere of HD~122563 calculated with the \COBOLD\ code. We then make a detailed comparison of the theoretical line shifts and bisectors with those derived from the high quality HARPS spectrum of HD~122563 taken from the ESO Science Archive Facility ($R=115\,000$, average signal-to-noise ratio, S/N $\approx$ 310). In general, we find a good agreement between the theoretically predicted and observed Doppler shifts of \ion{Fe}{i} line cores, with somewhat larger discrepancies seen in the case of weaker (equivalent width $W<5$\,pm) and stronger lines ($W>11$\,pm). Both observed and theoretical coreshifts cover a range between 0 and -1\,km/s, with increasingly stronger blueshifts for weaker lines and slight hints of a coreshift dependence on wavelength. Theoretical bisectors reproduce the observed ones reasonably well too, however, theoretical bisectors of the weak red ($\lambda > 600$\,nm) \ion{Fe}{i} lines have blueshifts that are by up to $\sim200$\,m/s larger than observed. The obtained results therefore suggest that the current \COBOLD\ models are capable of reproducing the large-scale velocity fields in the atmosphere of HD~122563 sufficiently well. Nevertheless, further efforts are needed in order to understand the physical reasons behind the discrepancies in theoretical predictions and observed properties of the weakest and strongest \ion{Fe}{i} lines.
    }

    \maketitle{}


    \section{Introduction}

Convection is an important physical phenomenon responsible for the energy transport in stellar atmospheres and interiors. In cool stars (such as the Sun), convection zone may reach close to (or even into) the atmospheric layers where the formation of  spectral lines takes place, and therefore may directly influence line strengths, shapes, and Doppler shifts. A direct imprint of convection is the granulation pattern seen on the surface of the Sun, comprised of hot rising granules and cooler intergranular downflows. Granulation cannot be directly resolved in other stars yet, with the exception of the closest supergiants such as Betelgeuse \citep[e.g.,][]{kervella09}, therefore it can be only detected indirectly, for example, through the line asymmetries and their Doppler shifts.

3D hydrodynamical stellar model atmospheres, such as those calculated with the \Stagger\ \citep{stein98} and \COBOLD\ \citep{freytag12} codes, treat convection in considerably more realistic way than the classical 1D models do. Up to now, \Stagger\ and \COBOLD\ models were successfully applied to study spectral line properties in the Sun \citep{asplund00, caffau11}, Procyon \citep{prieto02}, a main sequence turn-off star HD 74000 \citep{cayrel07}, several M dwarfs \citep{ramirez09}, and red giants \citep{collet07, ramirez10, kucinskas13, dobrovolskas13}. These studies have shown that theoretical predictions were reasonably consistent with the observations, suggesting that current 3D hydrodynamical model atmospheres are already sufficiently realistic to account for the convective flows taking place in real stellar atmospheres.

In this work we present first results of our further application of the 3D hydrodynamical \COBOLD\ model atmospheres to study convection in red giant stars. We focus our investigation on the asymmetries and Doppler shifts of \ion{Fe}{i} lines in the atmosphere of the metal-poor giant HD~122563. Atmospheric parameters of this star have been re-derived recently by using {\tt HIPPARCOS} parallaxes to obtain its surface gravity, $\log g=1.6$ \citep{leeuven07}, and {\tt CHARA} interferometry to determine its effective temperature, $\Teff=4600$\,K \citep{creevey12}. Spectral line coreshifts and bisectors in HD~122563 have been studied recently with the aid of \Stagger\ models \citep{ramirez10}. All this makes HD~122563 an interesting target for a study with \COBOLD\ models, allowing to test the realism of the \COBOLD\ model atmospheres in the metal-poor giant regime, and to compare theoretical predictions obtained with the \Stagger\ and \COBOLD\ codes.

	\section{Observed spectrum of HD~122563}

In this study we used a publicly available reduced HARPS spectrum of HD~122563 extracted from the ESO Science Archive Facility\footnote{\url{http://archive.eso.org/eso/eso_archive_main.html}} (program ID 080.D-0347(A)). The spectrum was obtained in the wavelength range of $380-680$\,nm, with the average signal-to-noise ratio of S/N $\approx 310$ and spectral resolution of $R=115 000$. We used the pipeline-reduced spectrum which was processed with the HARPS Data Reduction Software\footnote{\url{http://archive.eso.org/archive/adp/ADP/HARPS/index.html}} (DRS). The continuum normalization was done using a second order polynomial fit with the {\tt Dech20T} spectral analysis package\footnote{\url{http://www.gazinur.com/DECH-software.html}}.

	\section{Three-dimensional \COBOLD\ model atmospheres and spectral line synthesis}

For the calculation of synthetic \ion{Fe}{i} lines we used two \COBOLD\ model atmospheres calculated with identical effective temperature and gravity (corresponding to those of HD~122563, $\Teff=4590$\,K and $\log g=1.6$, cgs), at two different metallicities, $\moh=-2.0$ and $-3.0$. An enhancement of 0.4\,dex was applied to the abundances of O, Ne, Mg, Si, S, Ar, Ca, and Ti, while for other elements solar-scaled abundances from \citet{grevesse98} were used \citep[solar-scaled abundances from][were used in case of C, N, and O]{asplund05}. The \COBOLD\ model atmospheres were calculated using a rectangular box of $200\times200\times170$ grid points and covering $4.2\times4.2\times2.1$\,Gm in physical dimensions ($x,y,z$, respectively). From the entire 3D model run, twenty 3D model snapshots (i.e., 3D model structures obtained at different instants in time) were chosen for the spectral synthesis of the \ion{Fe}{i} lines. The snapshots were selected in such a way as to ensure that properties of the 20 snapshot ensemble would match those of the entire 3D model run as closely as possible.

Synthesis of the \ion{Fe}{i} lines was performed under the assumption of local thermodynamic equilibrium (LTE), using for this purpose the \LINFOR\ spectral synthesis package\footnote{\url{http://www.aip.de/~mst/Linfor3D/linfor_3D_manual.pdf}}. Spectral lines were synthesized for different iron abundances to obtain sequences of \ion{Fe}{i} line profiles at $\moh=-2.0$ and $-3.0$, with the iron abundances in steps of 0.2\,dex. The obtained line profiles were then interpolated for the metallicity of HD~122563, $\feoh=-2.6$ \citep[the latter taken from][]{gratton96}. To speed up the line synthesis calculations, only every third point of the 3D model structure (in each horizontal direction) was used to synthesize the \ion{Fe}{i} lines. Tests made with the full 3D model structures showed that this simplification does not alter the resulting line bisectors by more than $\sim0.02$\,km/s.

For the line bisector and coreshift analysis (Sect.~\ref{sec:meas}) we used 81 non-blended \ion{Fe}{i} lines located in the wavelength range of $\sim$~400--670\,nm. Line wavelengths were taken from \citet{nave94}, the excitation potentials and line broadening parameters were extracted from the VALD database \citep{kupka99}. The rotational velocity (\vsini) was obtained by simultaneously fitting the line strength, Doppler shift, and \vsini\ of the theoretical and observed line profiles, keeping the instrumental broadening and continuum level fixed.

\begin{figure*}[]
\centering
\includegraphics[width=12 cm]{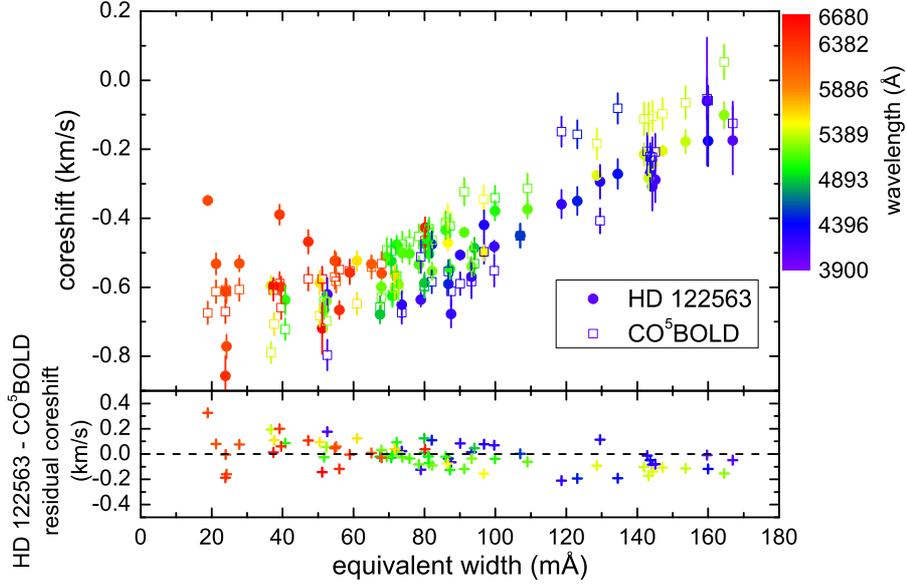}
\caption{Top panel: observed and theoretical coreshifts of \ion{Fe}{i} lines in the metal-poor red giant HD~122563, plotted versus line equivalent width, $W$. Errors of the observed coreshifts take into account  uncertainties in the laboratory wavelength, instrumental wavelength calibration, and photon noise. Error bars of the theoretical coreshifts show $1 \sigma$ of the coreshift distribution obtained from the Monte Carlo bootstrap experiment (see text for details). Bottom panel: difference between the observed and theoretical coreshifts (i.e., residual coreshifts) plotted versus $W$.}
\label{fig:coreshifts}
\end{figure*}

	\section{Line asymmetry measurements}
	\label{sec:meas}

Three spectral line properties were used to study the influence of convection on the formation of \fei\ lines: equivalent width ($W$, used as line strength indicator), line coreshifts (i.e., Doppler shifts of the deepest points in the line profiles), and line bisectors\footnote{The bisector is a line connecting loci of the midpoints between the blue and red wings of a spectral line profile at different flux (line depression) levels.}. Equivalent widths were measured by integrating theoretical line profiles obtained as best-fits to the given observed line profile. Both observed and theoretical coreshifts and bisectors were calculated using the following identical routines:
\begin{itemize}
	\item coreshifts were computed by fitting Gaussian profile to the five points centered around the deepest point in the observed/synthetic spectral line profile;
	\item bisectors were calculated utilizing the usual procedure of finding midpoints between the red and blue line wings at a number of different line depression levels. To obtain bisector values at any given flux value, a monotonized interpolation was used to interpolate between the observed/synthetic points on the red and blue wings of the line profile.
\end{itemize}

Errors in the observed line coreshifts were computed by adding uncertainties due to laboratory wavelengths, HARPS wavelength calibration, and photon noise (converted into wavelength uncertainty) in quadrature and taking a square root of the sum. We assumed that the error in the laboratory wavelength calibration was $2$\,m/s \citep[taken from][]{nave94} and the error in the wavelength calibration $\approx15$\,m/s (HARPS DRS estimate). The contribution from the photon noise to the coreshift error was estimated using the $\chi^2$ fitting procedure. For this, we assigned the measurement error to each flux point in the fitted line profile (calculated from the local $S/N$), fitted it with the Gaussian profile, and obtained the coreshift error corresponding to
the $\chi^2 \pm 1$ brackets. One should note that uncertainty due to the photon noise depends both on the line strength and wavelength ($S/N$ increases towards the red part of the spectrum).

In case of the observed line bisectors, we accounted only for errors due to photon noise which was computed using the formula of \citet{gray83}. Uncertainties due to the laboratory wavelength measurements and wavelength calibration errors were not taken into account since they do not change across the given spectral line profile.

It is important to remind that theoretical line profiles obtained using each of the twenty 3D model snapshot selection are in fact different from each other. Moreover, the properties of spectral lines computed using any different ensemble of the twenty 3D model snapshots should be slightly different from those obtained using the current snapshot selection. We therefore applied a bootstrap Monte Carlo method to estimate statistical uncertainty in the bisectors and coreshifts of the synthetic spectral lines. To this end, from the initial pool of twenty spectral lines computed using twenty 3D model snapshots we randomly selected twenty line profiles (allowing each profile to be selected several times). This procedure was repeated 1000 times, resulting in 1000 different sets of twenty theoretical line profiles. We then computed average core shifts and bisectors for each of these 1000 sets of spectral line profiles, and calculated their standard deviations (in the case of bisectors we computed standard deviations at a number of different flux levels). These standard deviations were then used as uncertainties in the theoretical line bisectors and coreshifts.

	\section{Results and discussion}
	\label{sec:results}

\begin{figure*}[]
\subfigure[]{\includegraphics[width=4.4 cm]{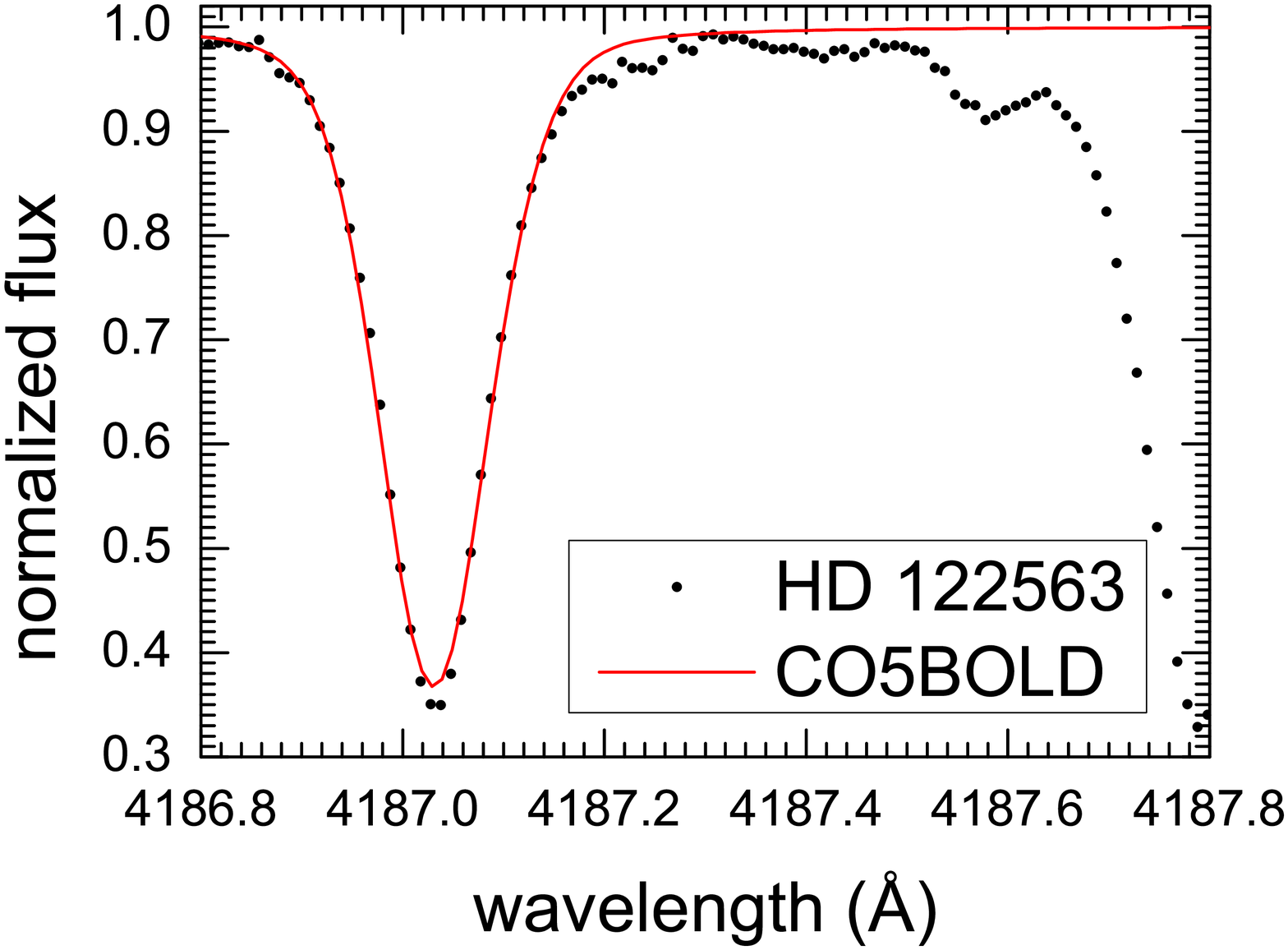}}
\subfigure[]{\includegraphics[width=4.4 cm]{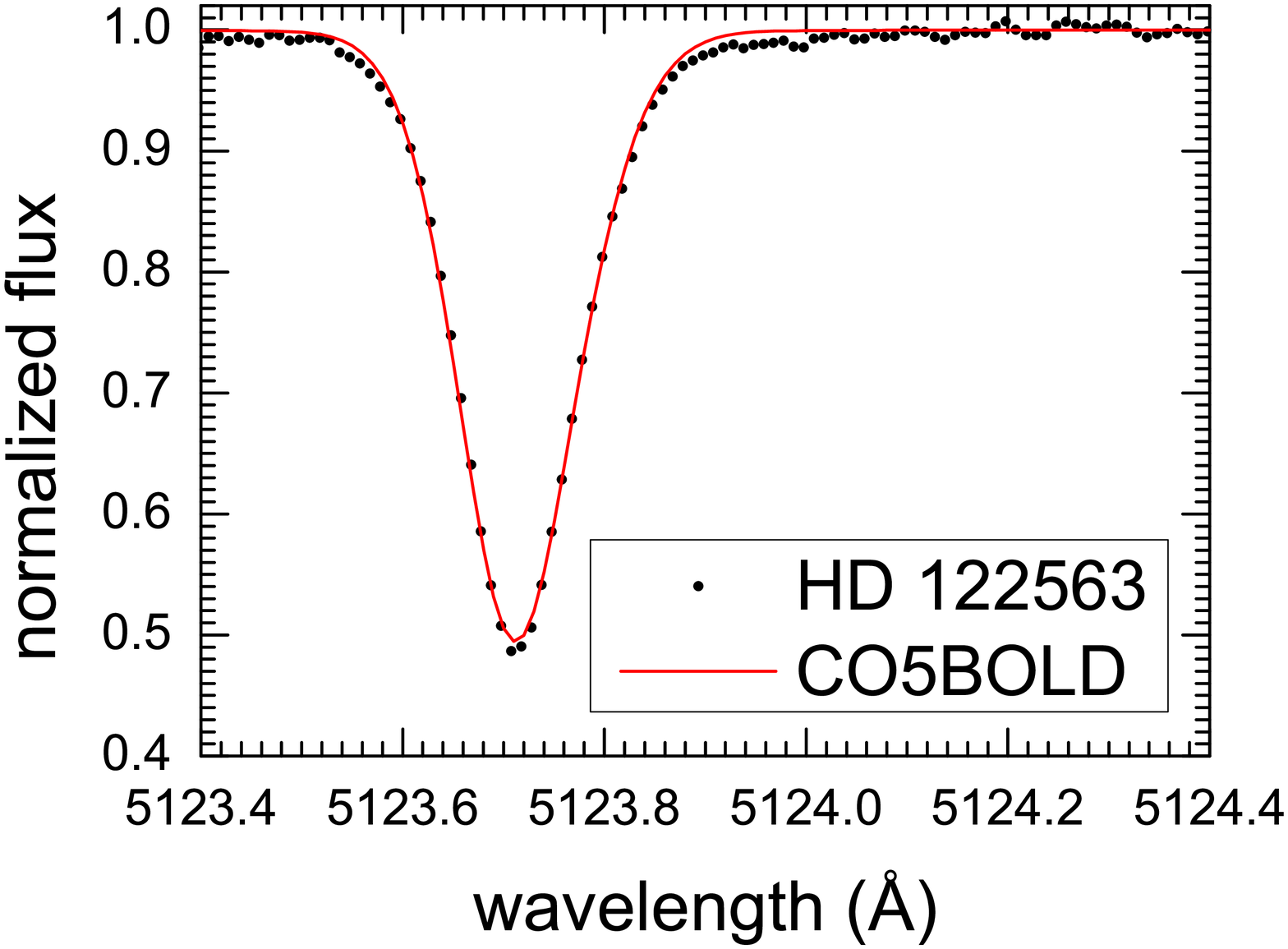}}
\subfigure[]{\includegraphics[width=4.4 cm]{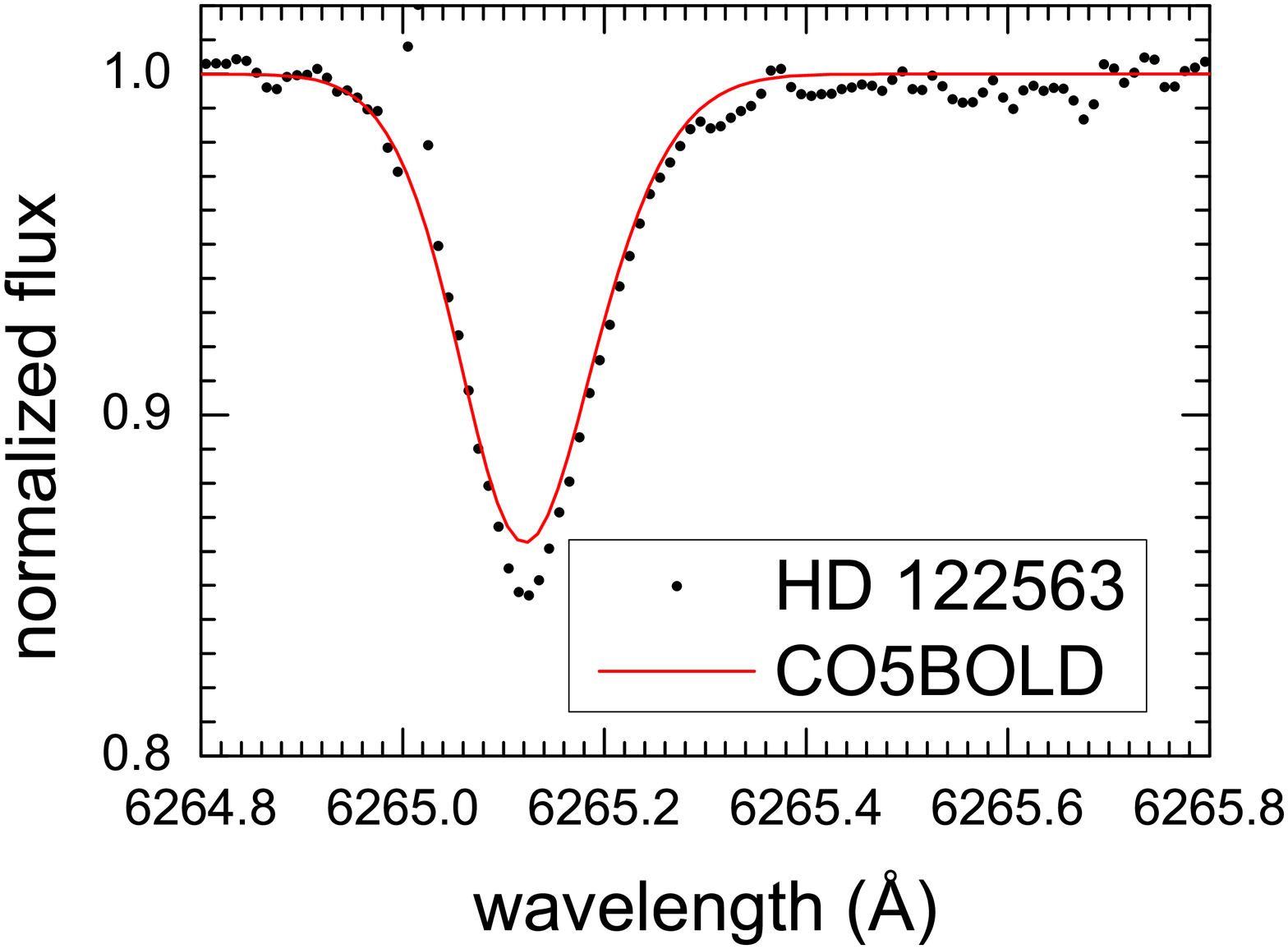}}\\
\subfigure[]{\includegraphics[width=4.4 cm]{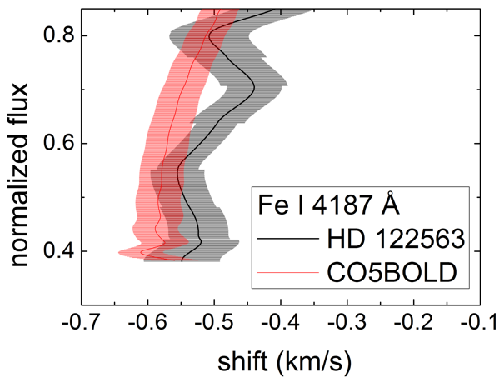}}
\subfigure[]{\includegraphics[width=4.4 cm]{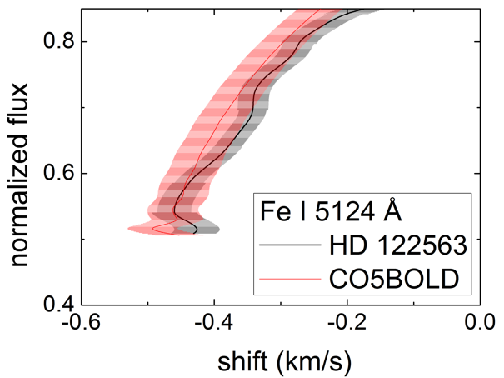}}
\subfigure[]{\includegraphics[width=4.4 cm]{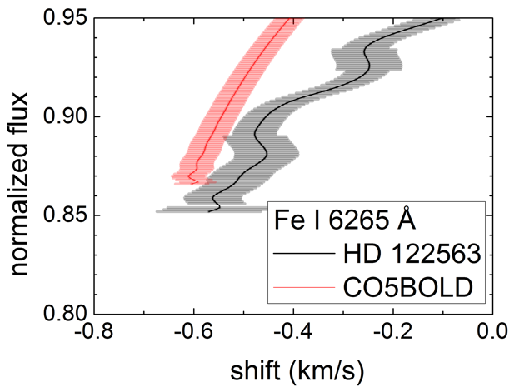}}
\caption{Top panels: examples of observed and best fitting theoretical \ion{Fe}{i} line profiles in the metal-poor red giant HD~122563. Bottom panels: bisectors of the spectral lines shown in the top panel. Gray shade represents the observed bisector uncertainty ($1\sigma$), red shaded region shows the $1\sigma$ uncertainty of the theoretical bisector.}
\label{fig:bis}
\end{figure*}

Before comparing observed and theoretical coreshifts, we determined the radial velocity ($v_{\rm rad}$) of HD~122563 by subtracting theoretical coreshifts from the observed ones. The mean residual was then added to the mean observed line shift to obtain the radial velocity, $v_{\rm rad}= -25.57\pm0.11$\,km/s (the error represents the standard deviation). An identical approach was utilized by \citet{ramirez10} who obtained a RV $v_{\rm rad}=-25.39\pm 0.09$\,km/s using 3D hydrodynamical \Stagger\ atmosphere models. The agreement between the two values obtained with different 3D hydrodynamical model atmospheres is indeed very good (the difference is 0.16 km/s).

The observed coreshifts (corrected for the radial velocity of HD~122563) generally agree well with those predicted using the \COBOLD\ models (Fig.~\ref{fig:coreshifts}). There are hints for the coreshift dependence on wavelength seen both in the observed and theoretical coreshifts, and these trends may be partly responsible for the scatter in the $W$-coreshift plane seen at any given value of $W$. The agreement between the observed and theoretical coreshifts is reasonably good in the case of medium strong lines ($5\,\mathrm{pm}<W<11\,\mathrm{pm}$), with the mean residual coreshift of $0.015\pm0.076$\,km/s (error is $1\sigma$ scatter around the mean value). The situation is slightly worse in the case of weakest ($W<5$\,pm) and strongest ($W>11$\,pm) lines, with the mean residual coreshifts of $+0.09\pm0.13$\,km/s and $-0.088\pm0.081$\,km/s, respectively. These discrepancies are about two times larger than the respective median error of the individual residual coreshift, 0.044\,km/s (weak lines) and 0.062\,km/s (strong lines).

Spectral line bisectors may be yet another tool to study velocity fields in stellar atmospheres, especially since they carry the imprints from the velocity fields and temperature fluctuations over the formation region of a given spectral line. In our study, observed and theoretical bisectors were compared by correcting theoretical bisectors for the radial velocity of HD~122563. We stress, however, that no additional corrections were applied to the theoretical bisectors, as it is sometimes done in other bisector studies \citep[e.g.,][]{ramirez10}. One may note that in case of both observed and theoretical bisectors there is a clear relation between the strength of a given spectral line and its bisector shape, with the bisectors of strong lines being nearly vertical and those of weaker lines becoming more curved, with the line cores getting progressively more blueshifted with decreasing line strength (Fig.~\ref{fig:bis}).

From the comparison of observed and theoretical bisectors provided in Fig.~\ref{fig:bis} we conclude that there is a reasonably good agreement between the observed and theoretical bisectors of strong lines and lines of intermediate strength (apart from the residual discrepancy in the coreshifts of the strong lines). However, in the case of weak lines discrepancies in both the coreshifts and bisector shapes are more pronounced. This result is similar to that obtained by \citet[][]{ramirez10} with the aid of \Stagger\ models, where the weakest theoretical bisector was blueshifted by 0.1\,km/s to match the observed bisector (the blueshifts required in case of stronger lines were smaller, at the level of $\sim0.03$\,km/s). Reasonably good agreement between theoretical predictions and observations may therefore suggest that currently available \COBOLD\ models are reproducing the global velocity fields in the atmosphere of HD~122563 sufficiently well. Nevertheless, further improvements may be needed in order to obtain better agreement in case of the weakest and strongest spectral lines.

\section{Conclusions}

Preliminary results obtained during the project aimed to study spectral line asymmetries in the metal-poor giant HD~122563 show that \ion{Fe}{i} line coreshifts and bisectors predicted by the current 3D hydrodynamical \COBOLD\ model atmospheres agree reasonably well with those observed in HD~122563, despite the minor problems in case of the weakest and strongest spectral lines. These results are compatible with those obtained in a similar study of line asymmetries in HD~122563 carried out using 3D hydrodynamical \Stagger\ model atmospheres, pointing to the consistency in the predictions made with two different 3D hydrodynamical model atmosphere codes. All this suggest that the current level of realism of the 3D hydrodynamical model atmospheres should be sufficient to warrant further applications of the 3D model atmospheres in a wider variety of astrophysical contexts, in order to further test their realism and identify problems where additional improvements may be needed.

\begin{acknowledgements}
We thank the organizers of the 2nd CO5BOLD workshop for the financial assistance helping JK, AK, and DP to attend the event. JK would like to thank M. Spite for providing the initial \ion{Fe}{i} list and E. Caffau for the discussions on the 3D model snapshot selection. This work was supported by grant from the Research Council of Lithuania (PRO-05/2012). HGL acknowledges financial support by the Sonderforschungsbereich SFB 881 ``The Milky Way System''� (subproject A4) of the German Research Foundation (DFG). AK and HGL acknowledge financial support from the the Sonderforschungsbereich SFB 881 ``The Milky Way System''� (subproject A4) of the German Research Foundation (DFG) that allowed  exchange visits between Vilnius and Heidelberg. PB and AK acknowledge support from the Scientific Council of the Observatoire de Paris and the Research Council of Lithuania (MOR-48/2011) that allowed exchange visits between Paris and Vilnius. MS acknowledges funding from the Research Council of Lithuania for a research visit to Vilnius.
\end{acknowledgements}

\bibliographystyle{aa}

\begin{thebibliography}{}
%\expandafter\ifx\csname natexlab\endcsname\relax\def\natexlab#1{#1}\fi

\bibitem[Allende Prieto et al.(2002)]{prieto02}
Allende Prieto,~C., Asplund,~M., Garc{\'{\i}}a L{\'o}pez,~R.~J., \& Lambert,~D.~L. 
2002, \apj, 567, 544

\bibitem[Asplund et al.(2000)]{asplund00} 
Asplund,~M., Nordlund,~{\AA}., Trampedach,~R., Allende Prieto,~C., \& Stein,~R.~F.
2000, \aap, 359, 729

\bibitem[Asplund et al.(2005)]{asplund05} 
Asplund, M., Grevesse,~N., \& Sauval, A.~J.
\ 2005, Cosmic Abundances as Records of Stellar Evolution and Nucleosynthesis, 336, 25

\bibitem[Caffau et al.(2011)]{caffau11}
Caffau,~E., Ludwig,~H.-G., Steffen,~M., Freytag, B., \& Bonifacio,~P.
2011, \solphys, 268, 255

\bibitem[Cayrel et al.(2007)]{cayrel07}
Cayrel,~R., Steffen,~M., Chand,~H., et al.
2007, \aap, 473, L37

\bibitem[Collet et al.(2007)]{collet07}
Collet,~R., Asplund,~M., \& Trampedach,~R.
2007, \aap, 469, 687

\bibitem[Creevey et al.(2012)]{creevey12}
Creevey,~O.~L., Th{\'e}venin,~F., Boyajian,~T.~S., et al.
2012, \aap, 545, A17

\bibitem[{Dobrovolskas} {et al.}(2013)]{dobrovolskas13}
Dobrovolskas,~V., Ku\v{c}inskas,~A., Steffen, M., Ludwig,~H.-G., Prakapavi\v{c}ius,~D., Klevas,~J., Caffau,~E., \& Bonifacio,~P.
2013, \aap, submitted

\bibitem[Freytag et al.(2012)]{freytag12}
Freytag,~B., Steffen,~M., Ludwig,~H.-G., et al.
2012, Journ. Comp. Phys., 231, 919

\bibitem[Gray(1983)]{gray83}
Gray,~D.~F.
1983, \pasp, 95, 252

\bibitem[Gratton et al.(1996)]{gratton96}
Gratton,~R.~G., Carretta,~E., \& Castelli,~F.
1996, \aap, 314, 191

\bibitem[Grevesse \& Sauval(1998)]{grevesse98}
Grevesse,~N., \& Sauval,~A.~J.
1998, \ssr, 85, 161

\bibitem[Kervella et al.(2009)]{kervella09}
Kervella,~P., Verhoelst,~T., Ridgway,~S.~T., et al.
2009, \aap, 504, 115

\bibitem[Ku{\v c}inskas et al.(2013)]{kucinskas13}
Ku{\v c}inskas,~A., Steffen,~M., Ludwig,~H.-G., et al.
2013, \aap, 549, A14

\bibitem[Kupka et al.(1999)]{kupka99}
Kupka,~F., Piskunov,~N., Ryabchikova,~T.~A., Stempels,~H.~C., \& Weiss,~W.~W.
1999, \aaps, 138, 119

\bibitem[van Leeuwen(2007)]{leeuven07}
van~Leeuwen,~F.
2007, \aap, 474, 653

\bibitem[Nave et al.(1994)]{nave94}
Nave,~G., Johansson,~S., Learner,~R.~C.~M., Thorne,~A.~P., \& Brault,~J.~W.
1994, \apjs, 94, 221

\bibitem[Ram{\'{\i}}rez et al.(2009)]{ramirez09}
Ram{\'{\i}}rez,~I., Allende Prieto,~C., Koesterke,~L., Lambert,~D.~L., \& Asplund,~M.
2009, \aap, 501, 1087

\bibitem[Ram{\'{\i}}rez et al.(2010)]{ramirez10}
Ram{\'{\i}}rez,~I., Collet,~R., Lambert,~D.~L., Allende Prieto,~C., \& Asplund, M.
2010, \apjl, 725, L223

\bibitem[Stein \& Nordlund(1998)]{stein98}
Stein,~R.~F., \& Nordlund,~{\AA}.
1998, \apj, 499, 914

\end{thebibliography}

\end{document}